\documentclass[conference]{IEEEtran}
\IEEEoverridecommandlockouts

\usepackage{cite}
\usepackage{amsmath,amssymb,amsfonts}
\usepackage[ruled,vlined]{algorithm2e}
\SetKw{KwTo}{to}
\SetKw{KwBy}{by}
\usepackage{float}
\usepackage{graphicx}
\usepackage{textcomp}
\usepackage{xcolor}
\usepackage{url}
\def\BibTeX{{\rm B\kern-.05em{\sc i\kern-.025em b}\kern-.08em
    T\kern-.1667em\lower.7ex\hbox{E}\kern-.125emX}}
\begin{document}

\title{A Concurrent Queue System for Multi-GPU Platforms: Application to Bellman-Ford SSSP\\
}

\author{\IEEEauthorblockN{Beyza Çavuşoğlu}
\IEEEauthorblockA{\textit{Computer Engineering} \\
\textit{Koc University}\\
Istanbul, Turkey \\
bcavusoglu23@ku.edu.tr  0009-0005-9120-3878}
}

\maketitle

\begin{abstract}
This paper presents the design
and implementation of a multi-GPU concurrent queue system using NVIDIA’s NVSHMEM.
The Bellman-Ford algorithm is used as a
case study to evaluate the performance of the proposed concurrent FIFO queue, with this multi-GPU implementation being the first known instance of its kind. 
Experimental results demonstrate that the multi-GPU queue implementation achieves a maximum speedup of 3.92x and an average speedup of 3.04x over the single-GPU baseline on four NVIDIA A100
GPUs. When applied to the Bellman-Ford Single-Source Shortest Path (SSSP) algorithm, the multi-GPU system achieves a
maximum speedup of 3.03× and an average speedup of 2.65× compared to the single-GPU implementation, tested on
10 graphs of different kinds
taken from the SuiteSparse Matrix Collection. 
\end{abstract}


\section{Introduction}
Efficient queue management is a fundamental challenge  in computing, especially in parallel and distributed systems where multiple threads or processes concurrently access shared data structures. Queues are widely used in applications such as task scheduling, event-driven simulations, and graph algorithms, and typically follow FIFO, LIFO, or priority-based policies depending on the application’s logic [12].

With the emergence of general-purpose computing on GPUs, there has been a growing demand not only to adapt these concurrent data structures to massively parallel architectures, but also to leverage GPUs as a necessary computational platform for their scalability. 
However, parallelizing queue operations on GPUs introduces challenges such as contention, irregular memory access, and synchronization overhead, especially in dynamic and irregular workloads. While there has been significant research on concurrent FIFO queues, implementations specifically designed for GPU architectures remain limited, as they often fail to fully exploit thread-level parallelism due to blocking behavior, inefficient atomic operations, and inadequate support for fine-grained concurrency [26], [6], [18]. To overcome these limitations, this paper adopts the Broker Queue (BQ) - a FIFO, concurrent, lock-free, and linearizable queue specifically designed for GPUs [15]. 

Despite their significant potential, multi-GPU systems introduce challenges related to synchronization, memory coherence, and communication between GPUs. One key solution to address these challenges is NVIDIA’s NVSHMEM  [20], which facilitates efficient communication and synchronization between GPUs in a multi-GPU setup, allowing for scalable and faster parallel processing. Unlike prior work, which has primarily focused on single-GPU or CPU-based queue implementations, this research explores and demonstrates the efficiency of concurrent queue operations in multi-GPU environments, filling a critical gap in the field [31,32, 33,34,36]. 
To achieve this, each GPU is assigned its own partition of the queue using a PGAS (Partitioned Global Address Space) model, where local ring buffers and metadata are shared across GPUs, enabling fully parallel enqueue and dequeue operations.

To evaluate the performance of the proposed multi-GPU concurrent queue system, the Bellman-Ford algorithm was selected as a benchmark application due to its naturally parallelizable structure and heavy reliance on queue operations. The Bellman-Ford algorithm, which solves the SSSP problem, is a fundamental algorithm in graph theory. The SSSP problem involves finding the shortest path from a given source vertex to all other vertices in a weighted graph, where the edge weights may be negative [4]. Bellman-Ford is particularly useful because it can handle graphs with negative weight edges, but its performance can become a bottleneck in large-scale graphs, particularly when implemented in parallel environments. Despite its importance, there has been limited research into the application of GPU systems for optimizing the Bellman-Ford algorithm, especially with respect to the concurrent queue systems that manage updates to the vertices during the algorithm’s execution [1, 5, 21, 35].

Overall, this paper makes the following contributions:
\begin{itemize}
\item \textbf{Multi-GPU Concurrent FIFO Queues:} A novel concurrent FIFO queue system optimized for multi-GPU environments, achieving a maximum speedup of 3.92× and an average speedup of 3.04× over single-GPU implementations on four NVIDIA A100 GPUs.
\item \textbf{Bellman-Ford SSSP Optimization:} The first implementation of the Bellman-Ford algorithm utilizing multi-GPU concurrent queues, demonstrating significant efficiency improvements. Testing on diverse graphs from the SuiteSparse Matrix Collection shows a maximum speedup of 3.03× and an average speedup of 2.65× compared to single-GPU implementations.
\end{itemize}

\section{Background and Related Work}

\subsection{Concurrent Queue Algorithms}
FIFO queues process items in the order they were inserted, making them well-suited for work distribution [11], [15]. These queues are categorized as blocking or non-blocking [17]. Blocking algorithms may cause delays if slower processes hinder faster ones, whereas non-blocking algorithms ensure at least one operation completes within bounded time, enhancing concurrency [23, 3, 13, 22, 24].

In multiprocessor systems, blocking algorithms suffer from performance drops due to scheduling delays, cache misses, or page faults. Non-blocking algorithms prevent such bottlenecks, however, they are not a single monolithic category but rather consist of three distinct types: lock-free, wait-free, and obstruction-free. Lock-free algorithms ensure system-wide progress but may allow individual thread starvation. Wait-free algorithms provide stronger guarantees by ensuring every thread completes in bounded steps [2]. Obstruction-free algorithms [14], are the weakest non-blocking type, guaranteeing progress only in the absence of contention. All wait-free algorithms are lock-free, and all lock-free algorithms are obstruction-free, but not vice versa.

\subsection{Linearizability and Fair Ordering}
These are crucial for ensuring predictable behavior in concurrent data structures. 
Linearizability ensures operations appear sequential, maintaining the order of invocation and response events. 
For example, if one dequeue occurs at \mbox{timestamp} 1 and another at timestamp 2, this order is maintained despite execution timing differences [15].

Fair ordering ensures equitable access to shared resources, preventing starvation. For example, if two operations are enqueued, the first one is always processed before the second, maintaining a predictable and orderly execution sequence. In FIFO queues, it guarantees that operations are processed in their enqueued order, ensuring efficiency and fairness in multi-threaded systems.

\subsection{GPU Queues}
GPUs achieve far higher overall throughput than CPUs by dedicating most of their transistor budget to massively parallel data processing, executing thousands of threads concurrently, whereas CPUs optimize for low-latency single-thread performance with larger caches and complex control logic [28]. CUDA threads access multiple memory spaces, per-thread local, shared, cluster shared, global, constant, and texture memory,each optimized for different performance characteristics and visibility levels. Shared and cluster-shared memory enable fast intra-block communication, while global, constant, and texture memory provide grid-wide access with persistence across kernel launches, supporting efficient data reuse and specialized read-only operations. 

While this memory hierarchy enables high intra-GPU parallelism, many real-world workloads require scaling beyond a single device, making efficient multi-GPU coordination essential. Strong scaling is often limited in traditional multi-GPU applications because CPU-driven communication in MPI or OpenSHMEM introduces synchronization overhead and GPU idling [29]. NVSHMEM overcomes these limitations by enabling GPU-initiated, one-sided communication within a Partitioned Global Address Space (PGAS) model, providing a unified memory abstraction across GPUs (Figure 1). This approach reduces CPU-GPU synchronization, supports long-running kernels that overlap computation and communication, and simplifies development by abstracting low-level communication details [20, 30]. Although variables reside in a shared PGAS, updates are initially local to each Processing Element (PE), as illustrated in Figure 1a; however, NVSHMEM’s put/get operations allow efficient remote access to these distributed values. Unlike MPI’s two-sided communication, which suffers from synchronization and protocol overheads, NVSHMEM’s one-sided primitives-mapped to RDMA or NVLink-enable low-latency data movement and fine-grained concurrency, offering superior scalability for large multi-GPU workloads.

While there has been research on concurrent FIFO queues, studies focusing specifically on GPU-based concurrent FIFO queues remain limited [26], [6], [18]. A key limitation is treating threads as isolated units rather than utilizing GPU warps, leading to high contention in CAS (Compare and Swap) operations [6]. The Combined and CAS Loop Queue (CCLQ) addressed this, achieving a 40× speedup over previous GPU-based approaches.

Selecting a FIFO, concurrent, lock-free, and linearizable queue is critical for GPU performance. FIFO ordering ensures that tasks are executed in the precise order
of their submission, while concurrency maximizes GPU resource utilization. Lock-free designs prevent bottlenecks, and linearizability guarantees atomic execution. These properties make the BQ an ideal choice for GPU implementation [15].
Unlike traditional lock-free queues that suffer from excessive failed CAS operations, BQ maintains high performance even under contention. Blocking queues, while stable, hinder load balancing in multi-queue setups. BQ overcomes these challenges, optimizing fine-granular parallel work distribution on GPU cores. Performance evaluations confirm BQ as the fastest queue in single-GPU environments [15], and extending it to multi-GPU systems is expected to yield even greater speedups.

\begin{figure}[htbp]
    \centering
    \includegraphics[width=\columnwidth]{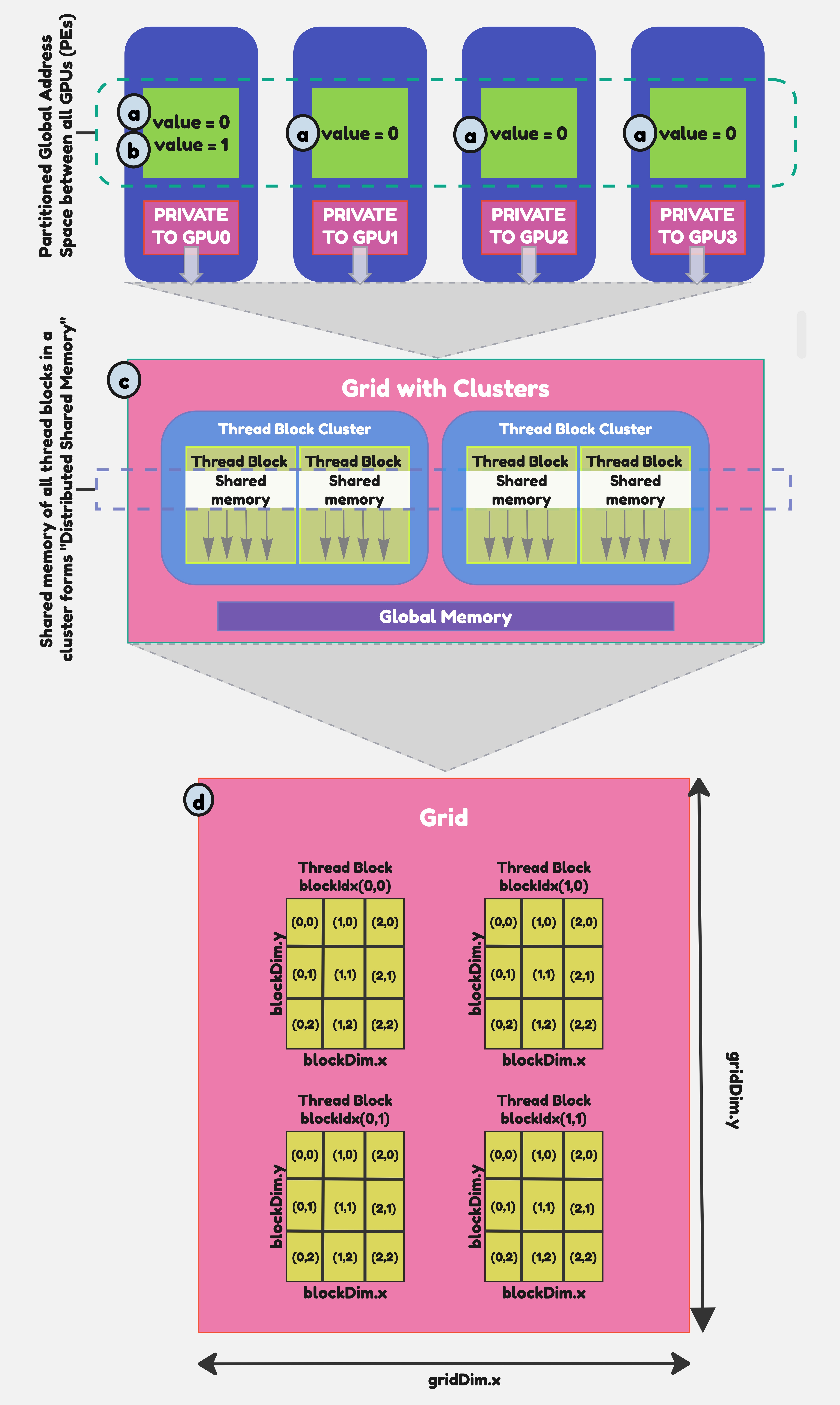}
    \caption{CUDA and NVSHMEM hierarchy a) value set to 0 in all PEs b) value set to 1 by PE1, only PE 1’s memory is updated, the rest of value variables in other PEs stayed as 0. c) GPU memory hierarchy with Thread Block cluster d) grid and block indexing in 2D}
    \label{fig:nvshmem_cuda}
\end{figure}

\subsection{SSSP  Algorithms - Bellman Ford}
The shortest-path problem finds the optimal route between two points with applications in routing, transportation, and social networks [16].
SSSP algorithms include BFS for unweighted graphs [9] and Dijkstra’s algorithm for non-negative weighted graphs, though the latter is hard to parallelize and cannot handle negative weights [16]. The Bellman-Ford algorithm [4],[10],[19] overcomes these issues by iteratively relaxing edges, supporting negative weights, and detecting cycles. While its time complexity is \( O(nm) \), it is more parallelizable [1,7]. While Dijkstra relies on a priority queue, Bellman-Ford can use simpler data structures, such as a FIFO queue [8]. 
This paper explores a parallel FIFO queue-based Bellman-Ford implementation for an efficient SSSP solution.

\section{Methodology and Implementation}
\subsection{Single-GPU Queue Implementation}
The BQ is a concurrent data structure for efficient enqueue and dequeue operations  [15]. It consists of a ring buffer, a fixed-size circular structure that overwrites old elements when full; head and tail pointers that track enqueue and dequeue positions; a ticket buffer that manages access sequencing to prevent conflicts; and a counter that maintains balance between operations for consistency. This algorithm ensures future operations by tracking count for linearizability and using a ticketing system to prevent race conditions and ensure fair ordering.
The pseudocode of the BQ algorithm is shown in Algorithm 
\ref{algo:broker}.

\subsubsection{Ensuring}
In traditional queue implementations, threads often skip enqueue operations if the queue is full, a limitation addressed by the broker mechanism in this algorithm. The broker ensures that operations are guaranteed before being executed. To do this, it uses a counter variable, "Count," that prevents modification of the head or tail pointers unless the operation can be completed successfully. For enqueue operations, the {\tt `ensureEnqueue`} method checks if there is space in the buffer or if items will soon be dequeued. Similarly, {\tt`ensureDequeue`} checks if an item is available for removal or if new items are being enqueued. This broker mechanism guarantees that only threads with a confirmed operation can modify the queue pointers, preserving the integrity of the queue in concurrent environments and ensuring linearizability.

\subsubsection{Ticketing}
To prevent race conditions and ensure orderly access, the algorithm implements a ticketing mechanism. Each ring buffer slot is associated with a ticket, and enqueue and dequeue operations are serialized based on the ticket number. Even-numbered tickets are for enqueue operations, while odd-numbered tickets are for dequeue operations. A thread must obtain a ticket for the slot it wishes to access and wait until its ticket matches the one in the buffer, ensuring controlled, sequential access to the queue and preventing race conditions. This mechanism guarantees fair ordering of operations and maintains the queue's integrity [15].

The "Broker Work Distributor Queue (BWD)" is an alternative to the Broker Queue, designed to simplify the enqueue and dequeue processes [15]. Instead of continuously checking for Full or Empty statuses, the {\tt `ensureEnqueue`} and {\tt`ensureDequeue`} functions are only called once during the enqueue and dequeue operations, providing a snapshot of the queue's state. While this design improves efficiency, it comes with a limitation: non-linearizability. The `Count` variable is used as a temporary assurance but does not always reflect the actual state of the queue. Despite this, the BWD is suitable for work distribution, as it ensures the queue eventually becomes empty unless new items are added. This behavior is sufficient for practical purposes in distributing tasks. 
Both the BQ and the BWD variants are used in the multi-GPU implementation, with the expectation that BWD will achieve better performance due to its reduced overhead from avoiding repeated state checks.

\begin{algorithm}[htbp]
\caption{Single-GPU Broker Queue Algorithm}
\label{algo:broker}
\SetKwFunction{FInit}{INITIALIZE}
\SetKwFunction{Fwaitticket}{WAIT\_FOR\_TICKET}
\SetKwFunction{FensureEnq}{ENSURE\_ENQUEUE}
\SetKwFunction{FputData}{PUT\_DATA}
\SetKwFunction{Fenqueue}{ENQUEUE}
\SetKwFunction{Freaddata}{READ\_DATA}
\SetKwFunction{FensureDeq}{ENSURE\_DEQUEUE}
\SetKwFunction{Fdequeue}{DEQUEUE}

\SetKwProg{Fn}{Function}{:}{}
\Fn{\FInit()} {
    $ head \gets 0 $; \\
    $ tail \gets 0 $; \\
    $ count \gets 0 $; \\
    $ Tickets[N] \gets \{0, 0, \ldots, 0\} $; \\
    $ RingBuffer[N] \gets \{0, 0, \ldots, 0\} $; \\
}

\SetKwProg{Fn}{Function}{:}{}
\Fn{\Fwaitticket(Pos, Expected)} {
    \While{$ Ticket[Pos] \neq \text{Expected} $}{
        \text{sleep()}\;
    }
}

\setlength{\columnsep}{10pt} 

\SetKwProg{Fn}{Function}{:}{}
\Fn{\FensureEnq()}{
    $ Num \gets \text{atomicLoad(count)} $; \\
    $ ensurance \gets \text{false} $; \\
    \While{$ \text{Num} < N \text{ and } \text{!ensurance} $}{
        \If{$ \text{atomicAdd(count,1)} < N $}{
            $ ensurance \gets \text{true} $\;
        }
        \Else{
            $Num\gets \text{atomicSub(count,1)}-1 $;\
        }
    }
    \KwRet{$ ensurance $;}
}

\SetKwProg{Fn}{Function}{:}{}
\Fn{\FensureDeq()}{
    $ Num \gets \text{atomicLoad(count)} $; \\
    $ ensurance \gets \text{false} $; \\
    \While{$ \text{Num} > 0 \text{ and } \text{!ensurance} $}{
        \If{$ \text{atomicAdd(count, -1)} <= 0 $}{
            $ \text{atomicAdd(count,1)} $;\\
            $ Num \gets \text{atomicLoad(count)} $;\
        }
    }
    \KwRet{$ ensurance $;}
}

\vspace{0.5cm}
\setlength{\columnsep}{10pt} 

\SetKwProg{Fn}{Function}{:}{}
\Fn{\FputData(data)}{
    $ Pos \gets \text{atomicAdd(tail,1)} $; \\
    $ P \gets \text{Pos \% N} $; \\
    $ \text{WAIT\_FOR\_TICKET(P, 2*(Pos/N))} $; \\
    $ RingBuffer[P] \gets \text{data} $; \\
    $ Tickets[P] \gets \text{2*(Pos/N) + 1} $; 
}

\SetKwProg{Fn}{Function}{:}{}
\Fn{\Freaddata(data)}{
    $ Pos \gets \text{atomicAdd(head,1)} $; \\
    $ P \gets \text{Pos \% N} $; \\
    $ \text{WAIT\_FOR\_TICKET(P, 2 * (Pos / N) + 1)} $; \\
    $ poppedData \gets \text{RingBuffer[P]} $; \\
    $ Tickets[P] \gets \text{2*(Pos + N/N)} $; 
}

\setlength{\columnsep}{10pt} 
\vspace{0.5cm}

\SetKwProg{Fn}{Function}{:}{}
\Fn{\Fenqueue(data)}{
    \While{not $ \text{ENSURE\_ENQUEUE()} $}{
        \If{$ N \leq (tail - head) < N + \text{MaxThreads}/2 $}{
            \KwRet{$ \text{Full} $;}
        }
    }
    \text{PUT\_DATA(data)};\\
    \KwRet{$ \text{Successful} $;}
    \vspace{0.2cm} 

}

\SetKwProg{Fn}{Function}{:}{}
\Fn{\Fdequeue(data)}{
    \While{not $ \text{ENSURE\_DEQUEUE()} $}{
        \If{$ N + \text{MaxThreads}/2 \leq (tail - head - 1) $}{
            \KwRet{$ \text{Empty} $;}
        }
    }
    \KwRet{$ \text{READ\_DATA(data)} $;}
}

\end{algorithm}

\subsection{Multi-GPU Queue Implementation}
Traditional single-GPU queue implementations typically rely on a centralized structure where all enqueue and dequeue operations are directed to a single ring buffer residing in that GPU’s memory. While straightforward, this centralized design becomes a bottleneck under heavy concurrency due to frequent contention on shared variables such as the head and tail pointers. As more threads attempt to simultaneously enqueue or dequeue, serialized access degrades performance, limiting throughput and increasing latency.

\begin{figure}[htbp]
\centering
\includegraphics[width=\columnwidth]{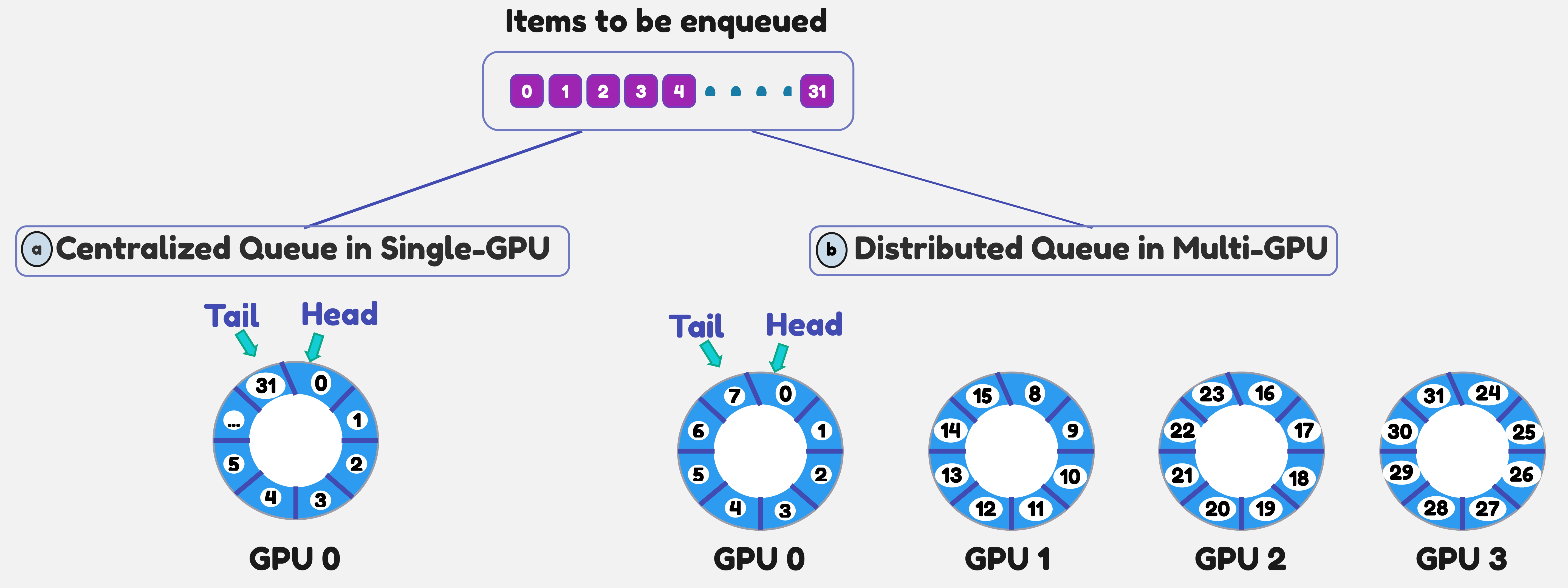}
\caption{Illustration of Single-GPU and Multi-GPU scenarios}
\label{fig:multi_conf}
\end{figure}

To overcome these limitations, multi-GPU systems distribute the workload and memory footprint across several GPUs. However, enabling efficient inter-GPU communication and coordination requires a careful memory management strategy. This paper presents a distributed queue implementation across multiple GPUs, where both data and operations are partitioned in a scalable fashion. Instead of centralizing the queue in one device, the global queue is logically divided among GPUs, with each GPU responsible for a local segment of the queue, managed independently but synchronized globally.

The centralized design (Figure 2a) places the entire queue within GPU 0, where all threads perform their operations on a single shared ring buffer. This leads to contention and potential performance degradation as the system scales. In contrast, the distributed design (Figure 2b) assigns each GPU a local queue segment with its own private ring buffer and ticket buffer. Each GPU executes enqueue and dequeue operations over its assigned portion of the global queue, reusing the single-GPU APIs with remote atomic updates to head and tail, significantly reducing contention and enhancing parallelism. The global queue is effectively distributed in a round-robin fashion across GPUs, enabling better utilization of memory and compute resources.

The implementation (Algorithm 2) is built upon the Partitioned Global Address Space (PGAS) model and uses NVSHMEM for efficient memory sharing and synchronization across GPUs. In this model, each PE, corresponding to a GPU, participates in a collective symmetric memory allocation. Specifically, NVSHMEM’s nvshmem\_malloc ensures that each GPU allocates identical-sized regions in its local symmetric heap, resulting in symmetric addresses that are valid and accessible across PEs. This enables any GPU to access another GPU’s memory location using a known symmetric address and remote PE ID [20].

In this design, each GPU’s queue segment (i.e., its portion of the ring buffer and ticket buffer) is logically private: it is primarily accessed by the GPU that owns it, as observed previously in Figure 1. However, due to symmetric allocation, these regions remain physically accessible by all other GPUs when needed for coordination or global queue management. In contrast, the coordination metadata-particularly the head and tail pointers used to track global enqueue and dequeue progress-is explicitly shared and accessed by all participating GPUs. This allows the system to maintain a coherent global view of the state of the queue even though the data are partitioned locally.

\begin{algorithm}[htbp]
\caption{Multi-GPU Broker/BWD Queue System}
\label{algo:mgpu_queue}
\SetKwFunction{Finit}{MultiGPU\_INIT}
\SetKwFunction{Fenq}{MultiGPU\_ENQUEUE}
\SetKwFunction{Fdeq}{MultiGPU\_DEQUEUE}
\SetKwFunction{FsingleEnq}{SINGLE\_GPU\_ENQUEUE}
\SetKwFunction{FsingleDeq}{SINGLE\_GPU\_DEQUEUE}


\SetKwProg{Fn}{Function}{:}{}
\Fn{\Finit()}{

    \tcp{Allocate queue in NVSHMEM PGAS instead of local GPU memory}
    $ head \gets$ nvshmem\_malloc(sizeof(int))\;
    $ tail \gets$ nvshmem\_malloc(sizeof(int))\;
    $ count \gets$ nvshmem\_malloc(sizeof(int))\;

    $ tickets[N] \gets$ nvshmem\_malloc(N)\;
    $ ring[N] \gets$ nvshmem\_malloc(N)\;

    \tcp{All GPUs initialize their local copy}
    \If{nvshmem\_my\_pe() == 0}{
        $head \gets 0$, $tail \gets 0$, $count \gets 0$\;
        \For{$i \gets 0$ \KwTo $N-1$}{
            $tickets[i] \gets 0$; \\
            $ring[i] \gets 0$;
        }
    }

    nvshmem\_barrier\_all()\;
}


\SetKwProg{Fn}{Function}{:}{}
\Fn{\Fenq{$value$}}{

    $pe \gets$ nvshmem\_my\_pe()\;
    $numGPUs \gets$ nvshmem\_n\_pes()\;

    \tcp{Compute partition range for this GPU}
    opsPerGPU $\gets numOperations / numGPUs$\;
    start $\gets pe \cdot$ opsPerGPU\;
    end   $\gets$ (pe $==$ numGPUs-1) ? $numOperations$ : start + opsPerGPU\;
    
    \For{$i \gets start + threadId$ \KwTo $end$ \KwBy $totalThreads$}{

        success $\gets$ \FsingleEnq(value,\,
                     pe,\,
                     count,\,
                     tail,\,
                     tickets,\,
                     ring)\;

        \If{success}{
            atomicAdd(succesfulEnqueues, 1)\;
        }
    }
}


\SetKwProg{Fn}{Function}{:}{}
\Fn{\Fdeq{}}{

    $pe \gets$ nvshmem\_my\_pe()\;
    opsPerGPU, start, end defined as above\;

    \For{$i \gets$ start + threadId \KwTo end \KwBy totalThreads}{

        data $\gets$ \FsingleDeq(count,\,
                                 pe,\,
                                 head,\,
                                 tickets,\,
                                 ring)\;

        \If{data $\neq$ EMPTY}{
            atomicAdd(succesfulDequeues, 1)\;
        }
    }
}

\end{algorithm}

\subsection{Single and Multi-GPU Bellman-Ford}

The Single-GPU Bellman-Ford implementation utilizes the aforementioned BQ structure that holds edges for relaxation, enabling concurrent enqueueing and improving efficiency. The algorithm begins by transferring the graph data—represented as Edge structures with source, destination, and weight fields—to GPU memory for parallel processing. The distance array is initialized with the source vertex distance set to zero and all other vertices set to infinity. The parallel relaxation phase employs multiple threads that concurrently dequeue edges from the BQ and attempt to update vertex distances if shorter paths are found. Each thread performs atomic operations to maintain consistency when multiple threads modify the same vertex's distance. A  change counter is used to track updates during iterations, serving both as a mechanism for negative cycle detection and as an early termination condition. It is important to note that this work does not aim to provide the most optimized single-GPU Bellman-Ford implementation, as achieved in prior studies [27], but rather to evaluate a queue-based design in order to analyze its performance trends in a multi-GPU environment.

The multi-GPU Bellman-Ford (Algorithm 3) partitions the graph and distance array across GPUs using BQ or BWD queues for efficient distributed processing. Each GPU maintains its own broker queue and updates distances for its assigned vertices, reducing the inter-GPU communication
latency and avoids bottlenecks from centralized data access. 
Moreover, to optimize performance, queue management benefits from simpler atomic operations (e.g., `atomicAdd`) instead of complex compare-and-swap mechanisms. One-sided NVSHMEM operations (`put/get`) enable inter-GPU distance updates when needed. A key feature is the implementation of efficient peer-to-peer communication channels between GPUs, allowing for direct memory transfers without CPU intervention. Additionally, the implementation includes a distributed negative cycle detection mechanism that coordinates across all GPUs to identify negative cycles spanning multiple devices.

\begin{algorithm}[htbp]
\caption{Multi-GPU Bellman-Ford Algorithm}
\KwIn{Graph \( G \) with vertices \( V \) and edges \( E \), source vertex \( s \),  \( N\_GPUs \)}
\KwOut{Shortest distances \( d \) from source vertex, detection of negative cycles}

\SetKwFunction{FInitializeDistances}{INITIALIZE\_DISTRIBUTED\_DISTANCES}
\SetKwFunction{FRelax}{PARALLEL\_RELAX\_EDGES}
\SetKwFunction{FBellman}{MULTI\_GPU\_BELLMANFORD}

\SetKwProg{Fn}{Function}{:}{}
\Fn{\FInitializeDistances(\(d\_dist\), \(V\), \(source\))}{
   
    \ForEach{vertex index \(i\) assigned to this thread (by striding over all vertices)}{
        \If{\(i = source\)}{
            Use NVSHMEM to set \(d\_dist[i]\) to \(0\) on this PE\;
        }
        \Else{
            Use NVSHMEM to set \(d\_dist[i]\) to \(\infty\) on this PE\;
        }
    }
}
\vspace{0.25cm}
\SetKwProg{Fn}{Function}{:}{}
\Fn{\FRelax(Graph \( G \), GPU\_ID)}{
   Partition the edge list so each GPU gets a contiguous chunk of edges and stores in its own BQ or BWD\;
    
   \ForEach{edge  assigned to this thread in its GPU's chunk}{
        
        $owner1 \gets edge.v1 \bmod N\_GPUs$\;
        
        $owner2 \gets edge.v2 \bmod N\_GPUs$\;
        
\shortstack[l]{
\(dist1 \gets \texttt{nvshmem\_atomic\_fetch}(\)\\
\(\hspace{1.5em}d\_dist[edge.v1], owner1)\)
}\

\shortstack[l]{
\(dist2 \gets \texttt{nvshmem\_atomic\_fetch}(\)\\
\(\hspace{1.5em}d\_dist[edge.v2], owner2)\)
}\

        \If{$dist1 \neq \infty$}{
           
            \If{$dist1 + edge.weight < dist2$}{
                $new\_dist \gets dist1 + edge.weight$\;
                
     \shortstack[l]{
\(\texttt{nvshmem\_put}(d\_dist[edge.v2],\)\\
\(\hspace{1.5em}new\_dist, owner2)\)
}\
                
                Atomically increment $changes$\;
            }
        }
    }
}

\SetKwProg{Fn}{Function}{:}{}
\Fn{\FBellman(Graph \( G \), Source \( s \), N)}{

        Allocate \( head, tail, count, ring\_buffer, tickets \) in PGAS using \texttt{nvshmem\_malloc}\;

        \FInitializeDistances{G, s, GPU\_ID}\;

    \( hasNegativeCycle \gets \text{false} \)\;
    
    \( changes \gets \) shared atomic counter\;

    \For{int \( i = 0 \) to \( V \)}{
       \FRelax{G, GPU\_ID}\;
        
        \( totalChanges \gets \) reduce\_sum(changes) across all GPUs\;
        
        \If{\( i == V-1 \land totalChanges > 0 \)}{
            \( hasNegativeCycle \gets \text{true} \)\;
        }
    }

 \tcp{Gather final distances from distributed memory}
    \For{vertex \( v = 0 \) to \( V-1 \)}{
        \( ownerPE \gets v \bmod N\_GPUs \)\;
        
        \( d[v] \gets \texttt{nvshmem\_get}(d\_dist[v], ownerPE) \)\;
    }
    
    \KwRet{\( d, hasNegativeCycle \)}
}
\end{algorithm}

\section{Experimental Results and Evaluation}
Tests were conducted on a workstation with four NVLink-connected NVIDIA A100-SXM4-40GB GPUs using CUDA 12.3. The first phase assesses the performance of the single-gpu queue in various configurations: thread counts (32-512), block counts (64, 108, 216), L2 cache set-aside ratios (0-75\%), and memory types (unified vs. global) to determine optimal configurations and establish a performance baseline for evaluating multi-GPU speedup. Multi-GPU tests focus on global memory without L2-cache set-aside, under varying thread and block counts. 
Kernel execution introduces overhead from initialization, and memory setup. To mitigate this, a warm-up phase executes kernels ten times before performance measurements. Morever, data transfer for enqueue operations is completed beforehand, ensuring that only enqueue and dequeue times are tested. For Single-GPU tests, 10 million operations are performed on the queue in two scenarios: \\ 
1) \textbf{Global Enqueue-Dequeue (EnqDeq)}: Sequential enqueue followed by dequeue operations, isolating their performance.  2) \textbf{Global Mixed}: Concurrent enqueue and dequeue, with odd-numbered blocks performing dequeue and even-numbered blocks enqueue, simulating real-world workloads.

The second phase evaluates practical performance using the Bellman-Ford. Tests compare single vs. multi-GPU implementations across multiple thread and block variations but only using global memory and no L2 cache set-aside, using 10 SuiteSparse graphs [25]. Speedup is discussed when using up to 4 GPUs.

\subsection{Single and Multi-GPU Queue Implementation Results}
Both queue designs, in both single and multi-GPU configurations demonstrate strong scalability as thread and block counts increase (Figure 3 and 4), particularly between 32 and 256 threads, showing significant reductions in total execution time which comes from the ability to support highly concurrent enqueue–dequeue operations, enabling efficient parallelism. However beyond 256 threads, performance gains begin to plateau, indicating that the system approaches its saturation point for concurrency. Additionally, performance under Unified Memory is less stable than Global Memory, especially under high-concurrency conditions, which is expected given that Unified Memory generally exhibits higher-latency data access patterns, less predictable and is still undergoing performance refinement by NVIDIA [28].

In mixed-workload scenarios, the Broker Queue maintains high efficiency, exhibiting decreasing execution times as concurrency increases and effectively balancing enqueue and dequeue operations. The BWD Queue also functions correctly; however, its throughput remains lower than the Broker’s under mixed conditions. This difference arises from BWD’s non-linearizable design, where ensureEnqueue and ensureDequeue are invoked only once per operation rather than being continuously evaluated. As the queue state changes rapidly when enqueues and dequeues interleave, this single-shot check may not fully capture the current contention level, occasionally resulting in unnecessary rejections even when capacity becomes available shortly after. In contrast, the Broker continuously re-validates these conditions, enabling it to adapt to dynamic state changes and sustain higher parallelism. Consequently, while both queues behave predictably and scale with increasing concurrency, the Broker achieves superior performance in mixed workloads, highlighting how synchronization design and state accuracy directly influence throughput under highly interleaved access patterns.

\begin{figure}[H]
    \centering
    \includegraphics[width=1\linewidth]{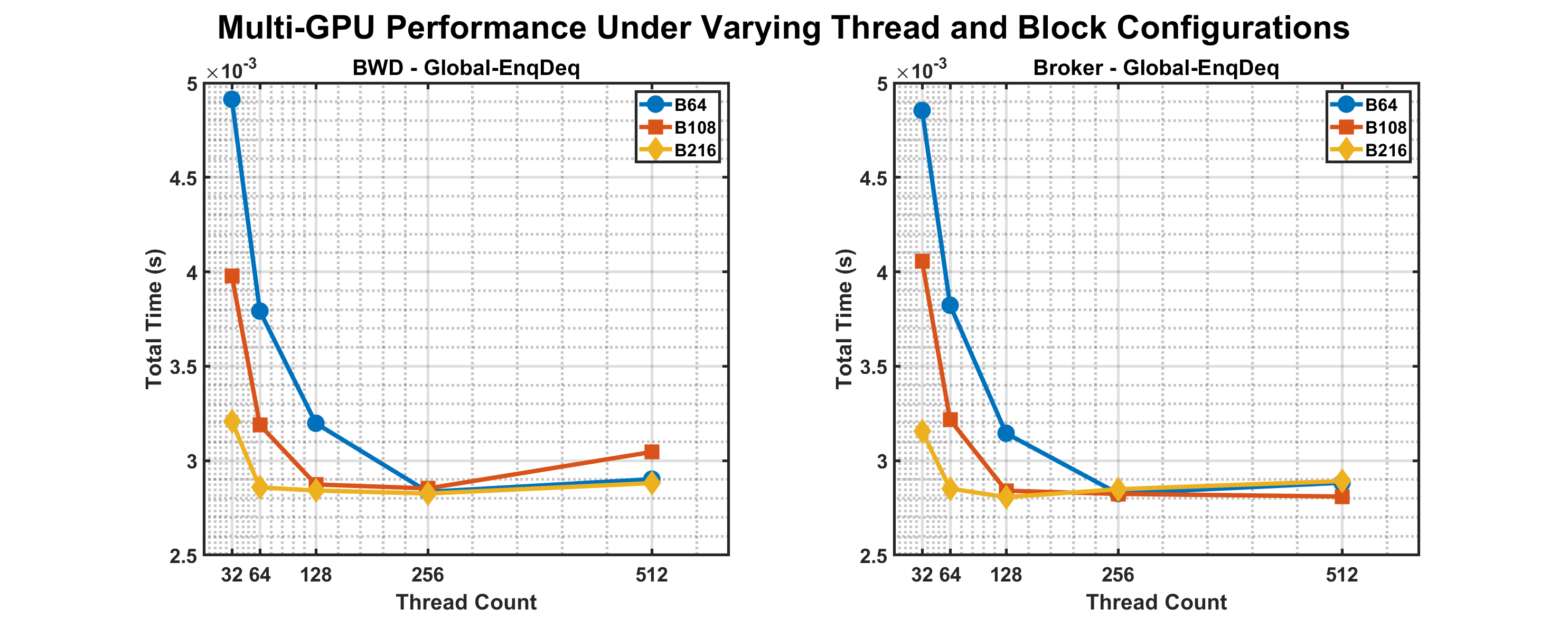}
    \caption{Multi-GPU Broker and BWD Queues under different thread and block counts}
    \label{fig:multigpu}
\end{figure}

\begin{figure}[htbp]
    \centering
       {\label{subfig:a}     \includegraphics[width=\columnwidth]{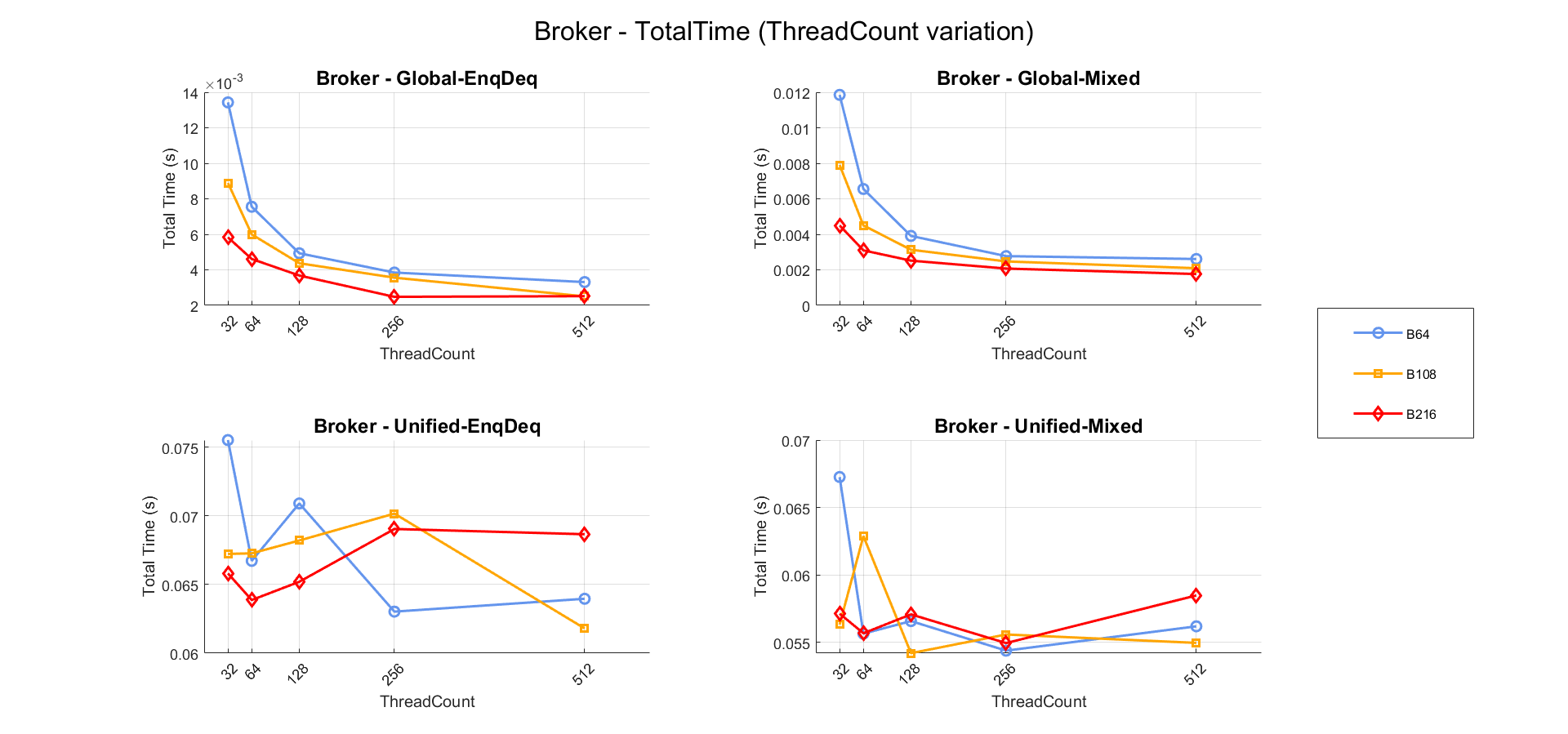}}\\
       {\label{subfig:b} \includegraphics[width=\columnwidth]{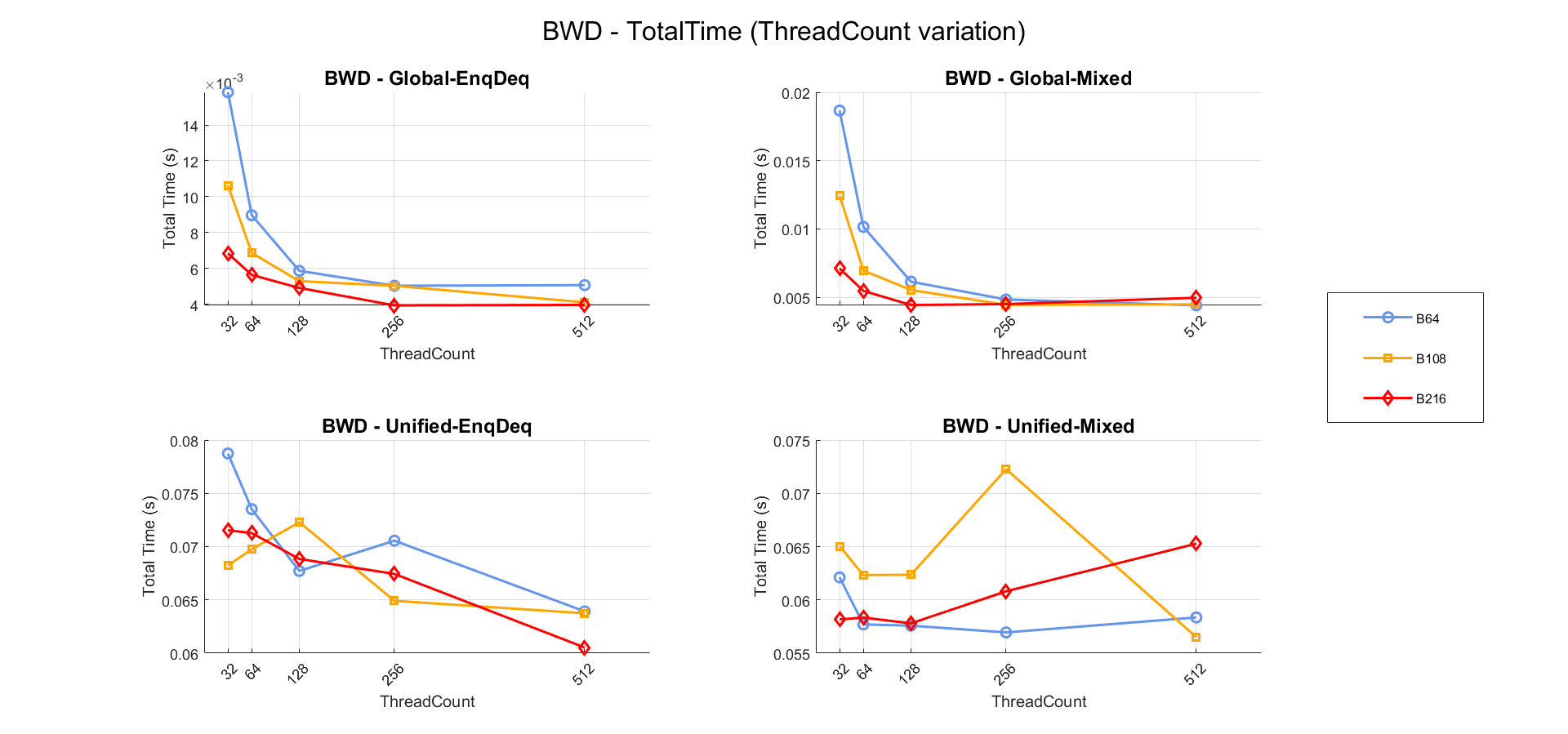}}
    \caption{(a) Single-GPU Broker Queue Unified and Global Memory under different thread and block counts, (b) Single-GPU BWD Queue Unified and Global Memory under different thread and block counts}
    \label{fig:sample}
\end{figure}

Figure 5 shows the speedup achieved by the Broker and BWD queues when transitioning from single-GPU to multi-GPU configurations. Both queues show diminishing speedup as thread counts increase, likely due to single-GPU efficiency and multi-GPU overhead. The maximum speedup for Broker Queue is 3.92x (average 3.04) and for BWD Queue, it’s 3.88 (average 3.02). With 2 GPUs, the maximum speedup is 1.42x (average 1.16), indicating performance may degrade with fewer GPUs.

\begin{figure}[H]
    \centering
     \includegraphics[width=0.85\linewidth]{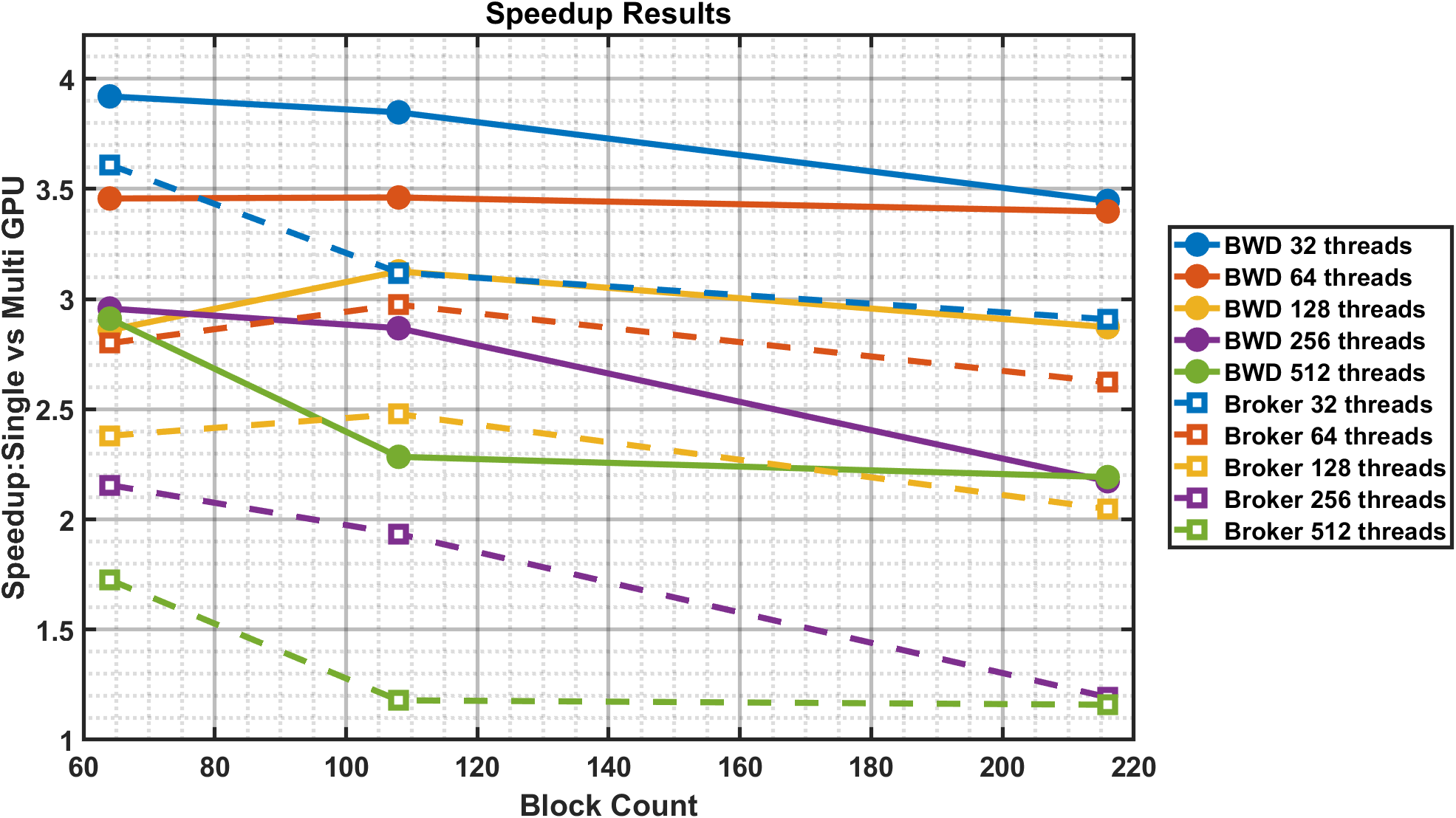}
    \caption{Speedup with 4 GPUs: Single-GPU vs Multi-GPU Broker and BWD Queues under different thread and block counts}
    \label{fig:speedup_mine}
\end{figure}

\subsection{Single and Multi-GPU Bellman-Ford Implementation Results} 
For single and multi-GPU implementations, both queues show an increase in execution time at higher thread counts, especially at 128 and 256 (Figures 6 and 7). Smaller block sizes (64) consistently show the lowest execution times, suggesting better resource utilization. However, as thread counts increase, performance diminishes due to lower occupancy, memory bandwidth saturation. Additionally, kernel launch overhead becomes significant when thread counts are high relative to block counts, diminishing parallelism benefits. Resource contention, especially in atomic operations, introduces serialization, further reducing efficiency. The Bellman-Ford algorithm's inherent dependencies also limit scalability, and as more threads are added, uneven workload distribution may lead to idle threads. 

\begin{figure}[H]
    \centering
    \includegraphics[width=1\linewidth]{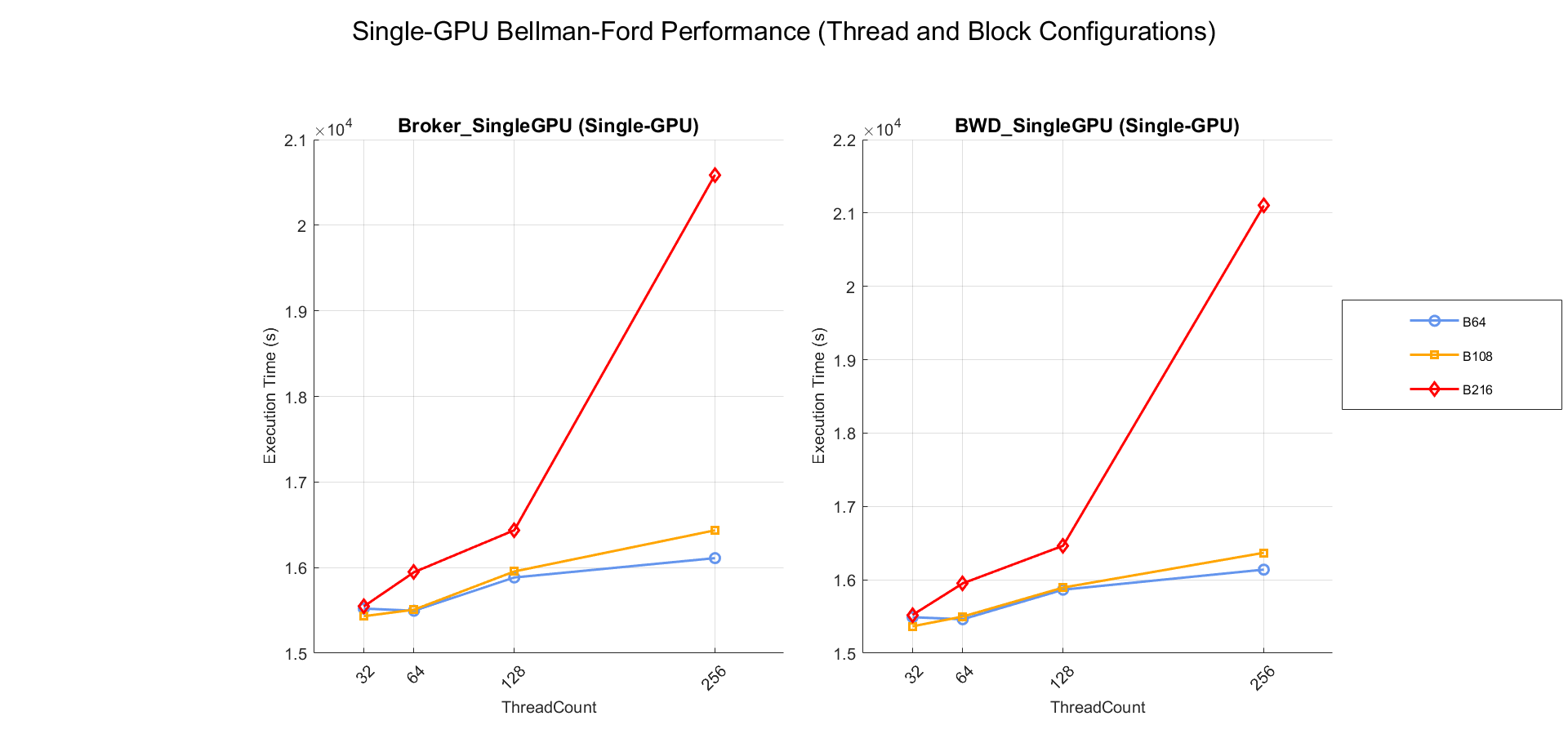} 
    \caption{Single-GPU BF for Broker and BWD Queues under different thread and block counts}
    \label{fig:single_bf}
\end{figure}

\begin{figure}[H]
    \centering
    \includegraphics[width=1\linewidth]{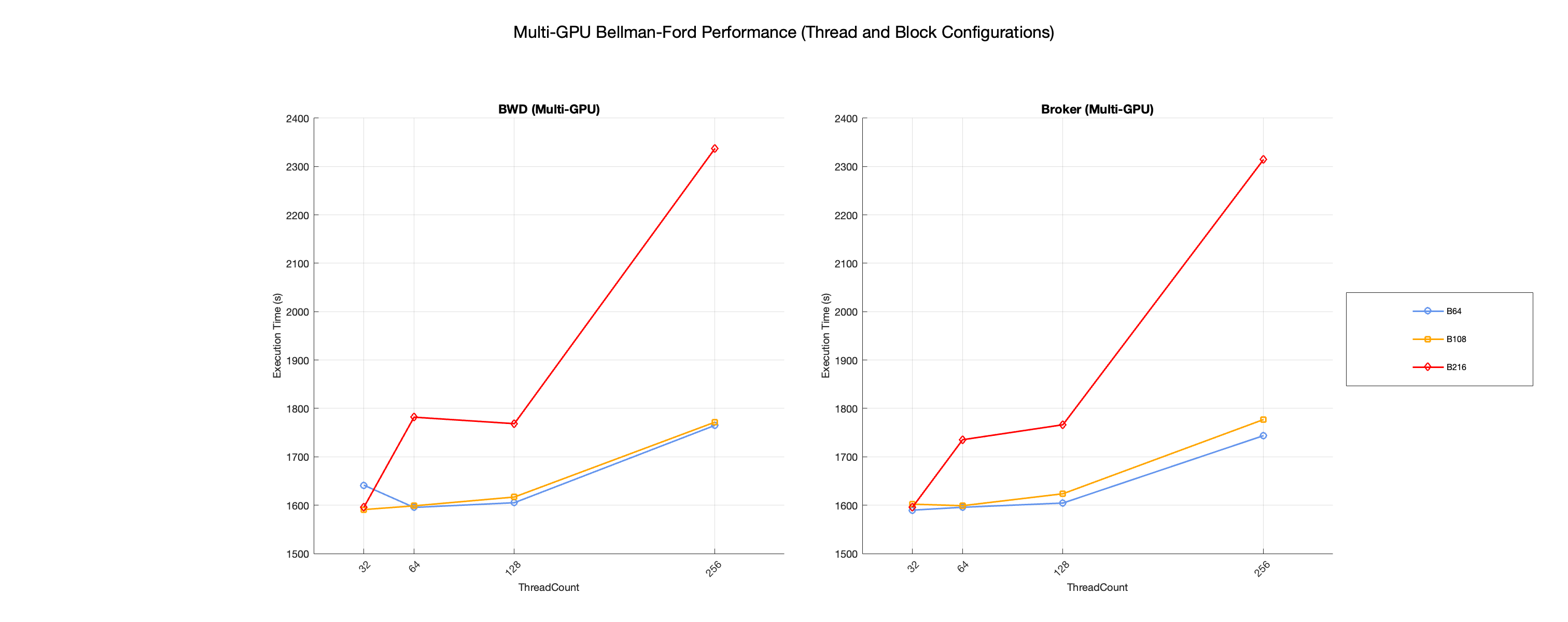}
    \caption{Multi-GPU BF for Broker and BWD Queues under different thread and block counts}
    \label{fig:multi_bf}
\end{figure}

\begin{figure}[H]
    \centering
\includegraphics[width=0.85\linewidth]{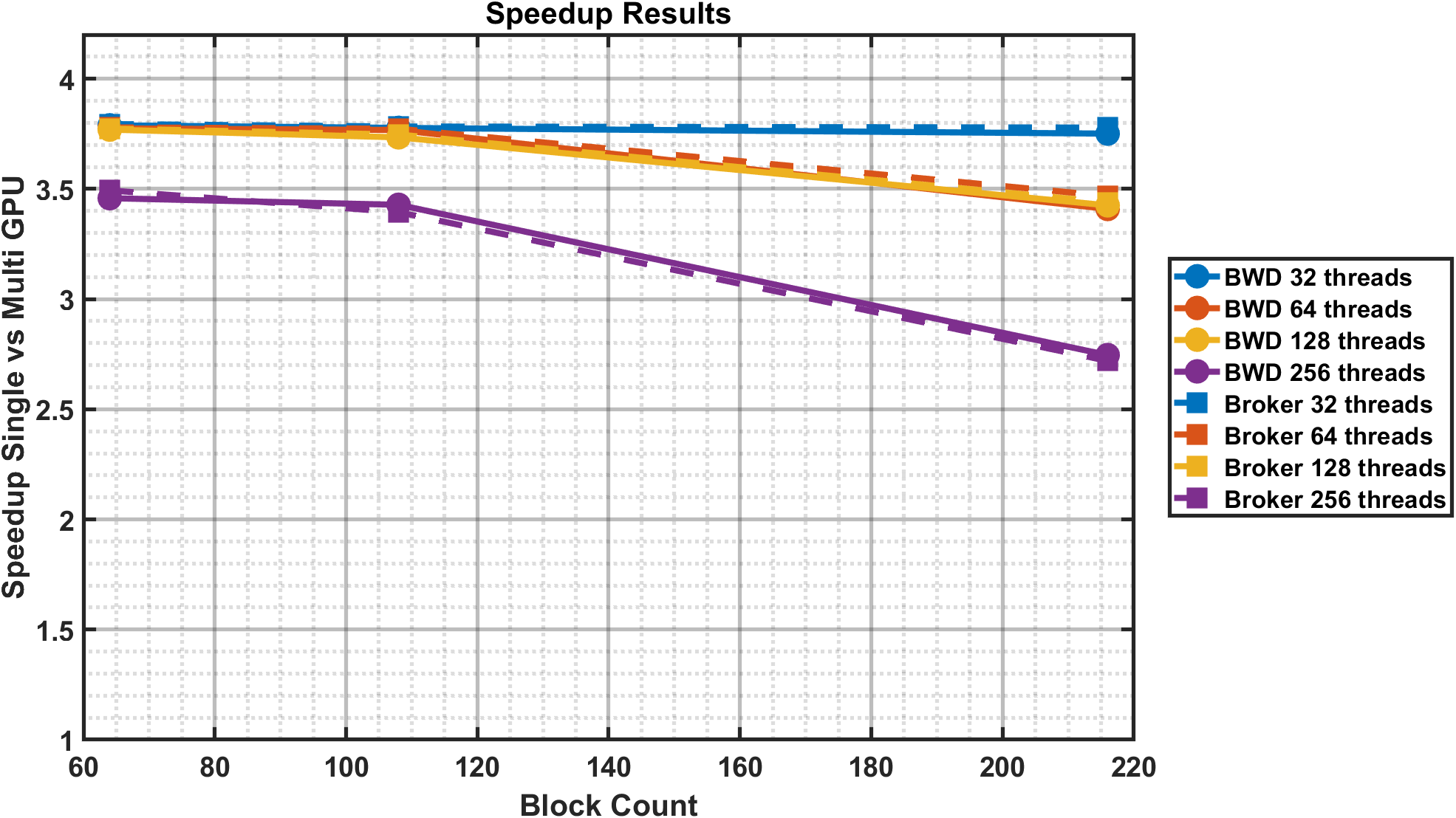} 
    \caption{Speedup over 4 GPUs: BF of Single-GPU vs Multi-GPU Broker and BWD Queues under different thread and block counts for SuiteSparse graph 'pcb3000'}
    \label{fig:speedup_bff}
\end{figure}

The speedup in Figure 8 compares the Broker and BWD queues for the Bellman-Ford algorithm with single and multi-GPU setups. Both queues show a decline in speedup as the block count increases, with better performance at 64 and 108 blocks, but worse at 216 due to overhead. For the 10 SuiteSparse graphs, the Broker queue achieves a maximum speedup of 3.03 and an average speedup of 2.65. Although Figure 5 shows speedups up to 3.92×, additional algorithmic overheads of Bellman-Ford might be limiting the gains.

Table 1 shows that with SuiteSparse graphs, the algorithm achieved important speedups, likely due to their complex connectivity, which improves data access patterns and reduces contention in the Bellman-Ford. The best result, from the pcb3000 graph (a linear programming problem), achieved a maximum speedup of 3.79× and an average of 3.57×, demonstrating the benefits of well-structured graph data.

\begin{table}[htbp]
\caption{SuiteSparse Graphs [25] and Single vs. Multi-GPU Speedup with 4 GPUs}
\centering
\scriptsize
\renewcommand{\arraystretch}{1.05}
\setlength{\tabcolsep}{2.5pt} 
\begin{tabular}{|c|c|c|c|c|c|}
\hline
\textbf{Graph} & \textbf{Kind} & \textbf{Vertices} & \textbf{Edges} &
\shortstack{\textbf{Max}\\\textbf{Speedup}} &
\shortstack{\textbf{Avg.}\\\textbf{Speedup}} \\ 
\hline
pcb3000 & Linear prog. & 3960 & 7732 & 3.79 & 3.57 \\ 
\hline
G65 & Random matrix & 8000 & 8000 & 3.70 & 3.48 \\ 
\hline
G67 & Random matrix & 10000 & 10000 & 3.18 & 3.01 \\ 
\hline
lp\_ken\_11 & Linear prog. & 14694 & 21349 & 3.10 & 2.96 \\ 
\hline
ch7-6-b5 & Combinatorial prob. & 5040 & 15120 & 2.99 & 2.45 \\ 
\hline
Alemdar & 2D/3D prob. & 6245 & 6245 & 2.32 & 2.21 \\ 
\hline
Reuters911 & Network & 13332 & 13332 & 2.32 & 2.23 \\ 
\hline
Foldoc & Net., computing & 13356 & 13356 & 2.32 & 2.23 \\ 
\hline
Cyl6 & Structural prob. & 13681 & 13681 & 2.31 & 2.23 \\ 
\hline
G39 & Random matrix & 2000 & 2000 & 2.29 & 2.14 \\ 
\hline
\textit{For all graphs} & -- & -- & -- & \textbf{3.03} & \textbf{2.65} \\ 
\hline
\end{tabular}
\label{tab:speedup}
\end{table}

\section{Conclusion}
This paper presents a multi-GPU concurrent FIFO queue system for graph processing workloads, extending single-GPU structures to improve efficiency in parallel processing. Performance evaluations show a maximum speedup of 3.92× and an average of 3.04× on four NVIDIA A100 GPUs, enhancing throughput and reducing latency.  
The Bellman-Ford algorithm, implemented with this novel queue, represents the first exploration of optimizing this algorithm using concurrent queues in a multi-GPU context. It achieves a maximum speedup of 3.03× and an average of 2.65×, demonstrating its effectiveness for graph processing.

\vspace{12pt}

\end{document}